\documentclass[a4paper,11pt]{article}
\usepackage{pos}

\title{Status of NNLO QCD corrections for process with one or more jets in the final state at the LHC}

\author*[a,b]{Jo\~{a}o Pires}

\affiliation[a]{LIP, Avenida Professor Gama Pinto 2, P-1649-003, Lisbon, Portugal}

\affiliation[b]{Faculdade de Ciências, Universidade de Lisboa, 1749-016 Lisboa}

\emailAdd{jnpires@lip.pt}

\abstract{
The abundant amount of data to be collected by the ATLAS and CMS collaborations in future runs of the Large Hadron Collider at CERN opens up a new era of precision physics. Some of the most prominent precision 
observables are related to processes with one or more jets in the final state. In order to fully exploit the potential of the LHC and the HL-LHC, it is imperative to make theoretical predictions at the level of accuracy that 
matches or even exceeds that of the upcoming measurements. In this talk we present a review of the status of theoretical predictions including NNLO QCD corrections for process with one or more jets in the final state at the LHC.
}

\FullConference{%
  *** Particles and Nuclei International Conference - PANIC2021 ***\\
  *** 5 - 10 September, 2021 ***\\
  *** Online ***
}

\begin{document}
\maketitle
\begin{figure}[!b]
\centering
\includegraphics[scale=0.3]{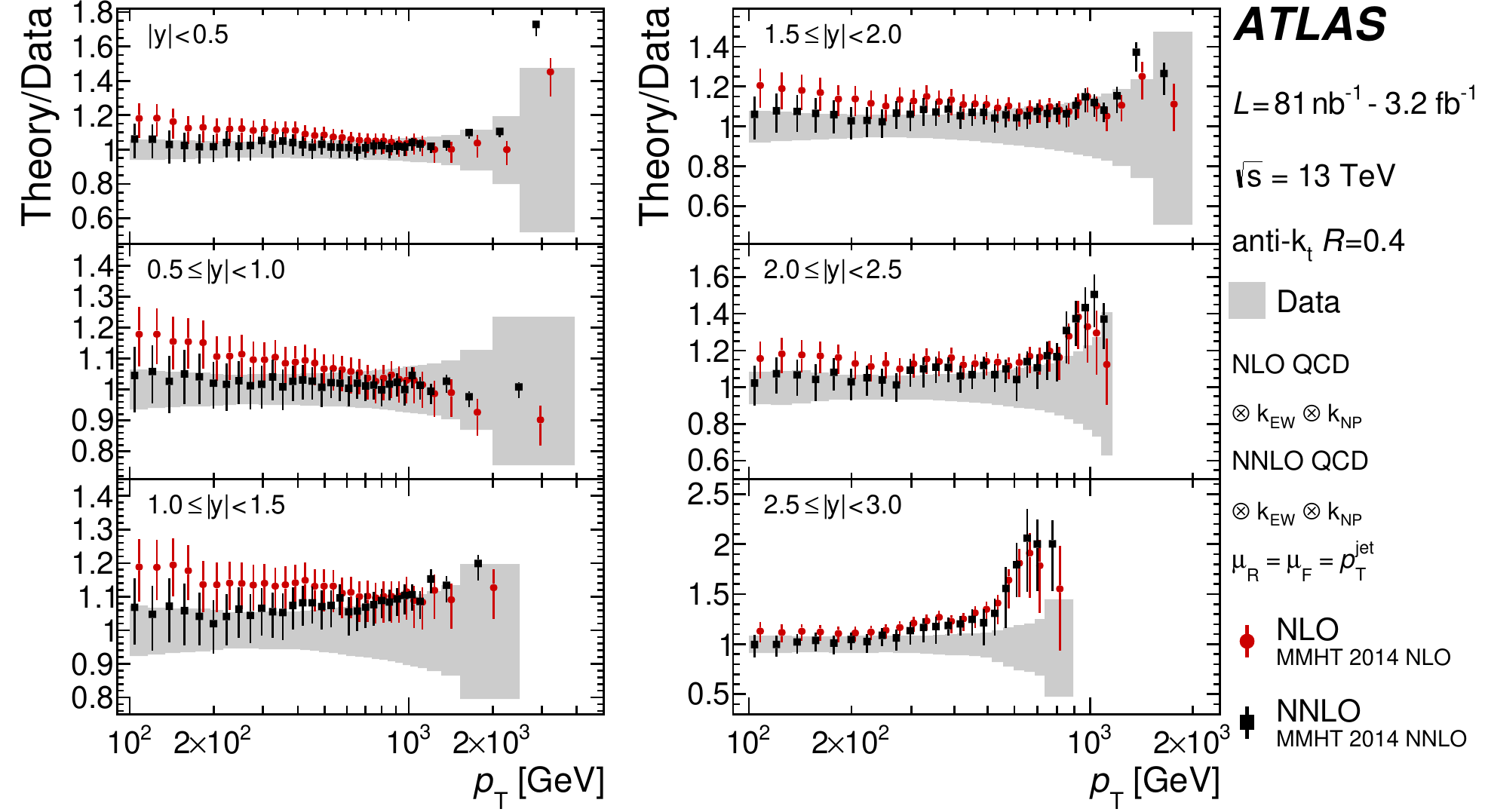}
\includegraphics[scale=0.45]{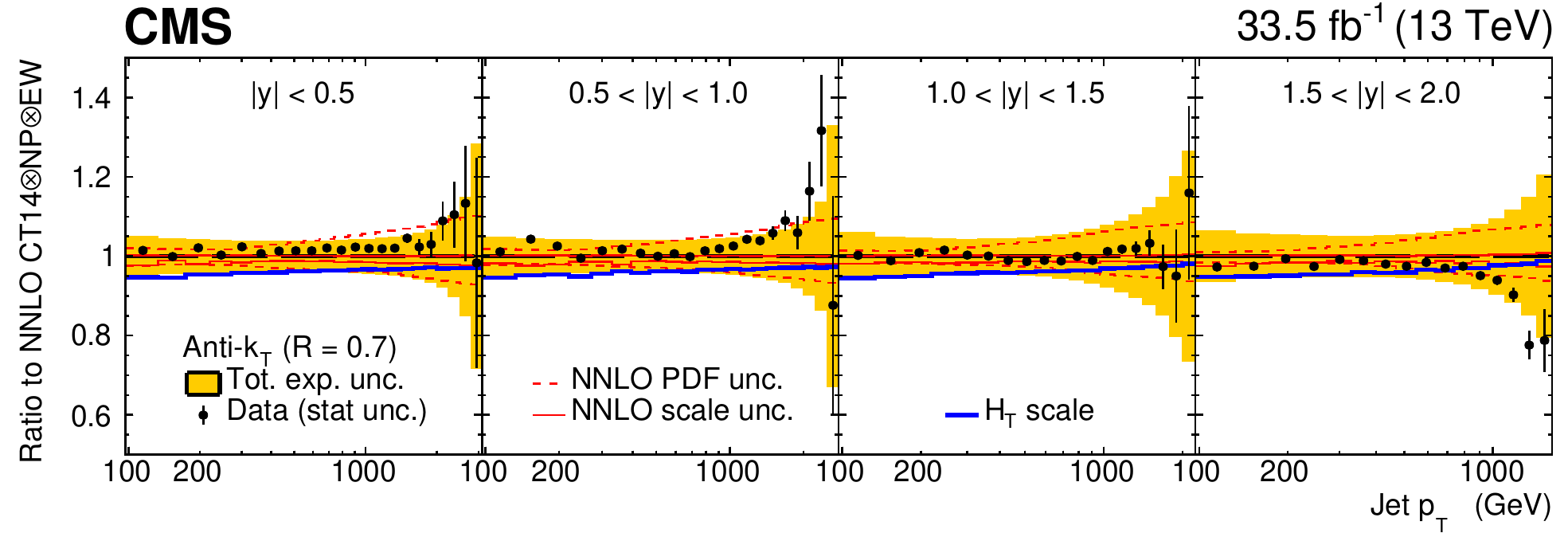}\\
\caption{Ratios of NNLO pQCD predictions and $\sqrt{s}$=13 TeV LHC measurements from ATLAS~\cite{ATLAS:2017ble} (left) and CMS~\cite{CMS:2021yzl} (right) for inclusive jet production.}
\label{fig:incjets}
\end{figure}

\section{Introduction}
The Large Hadron Collider (LHC) is currently colliding protons at centre of mass energies up to $\sqrt{s}=$ 13 TeV, with the goal of searching the high energy frontier for signs of physics beyond the Standard Model. To this end, it is mandatory to 
understand well the Standard Model and the collider environment of the LHC itself. In particular, the experimental precision reached by the LHC experimental collaborations currently requires theoretical predictions beyond the NLO in the perturbative QCD
expansion. It is for this reason that the inclusion of NNLO QCD corrections is today crucial to make reliable comparisons between data and theory without being limited by the uncertainty of the theoretical calculations. 
In this presentation we review the status of the NNLO QCD corrections for processes with one or more jets in the final state at the LHC. As we will see, these calculations lead to a great improvement in the predictions 
and to a decrease in the theoretical uncertainties. 

\section{Inclusive jet production}
The perturbative calculations for the inclusive jet and dijet cross sections at NNLO in QCD have recently been completed by two groups~\cite{Currie:2016bfm,Currie:2017eqf,Gehrmann-DeRidder:2019ibf,Czakon:2019tmo}. A comparison 
between the two independent calculations for inclusive jet production at $\sqrt{s}=$ 13 TeV in hadron-hadron collisions shows good agreement at NNLO and that subleading colour effects at this order in perturbation theory are negligible~\cite{Czakon:2019tmo}. These results are an important input for Parton Distribution Function (PDF) and strong coupling constant fits to hadron-collider data. In Fig.~\ref{fig:incjets} we show comparisons between fixed-order NLO and NNLO QCD predictions to ATLAS (left) and CMS (right) 13 TeV inclusive jet-$p_T$ spectrum data for anti-$k_{T}$ $R=0.4$ (ATLAS) and $R=0.7$ (CMS) jet cone sizes, corrected for non-perturbative and electroweak effects. We observe a significant improvement in the description of the jet data when going from NLO to NNLO and a smaller impact of the higher order corrections for the $R=0.7$ jet cone size. 

\begin{figure}[!t]
\centering
\includegraphics[scale=0.3]{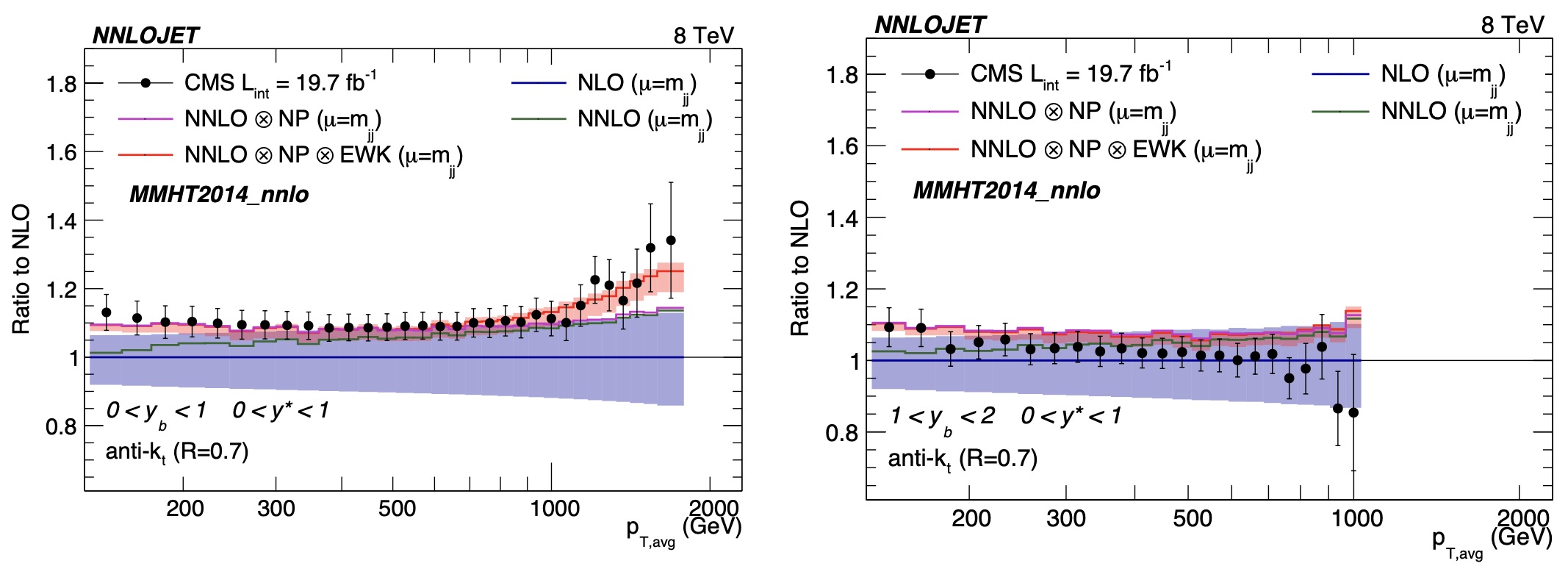}
\caption{Ratios of NLO and and NNLO theory predictions and CMS data normalized to the NLO central value.  The shaded bands shown for the NLO (blue) and the NNLO corrected for non-perturbative and electroweak effects (red) 
represent the theory uncertainty from variation of the renormalization and factorization scales in the calculation~\cite{Gehrmann-DeRidder:2019ibf}.}
\label{fig:dijets}
\end{figure}

For inclusive dijet data we consider at least two jets recorded by the detector and define the kinematical observables with the two jets with the highest transverse momentum. A calculation of dijet production at NNLO in QCD has been
presented in~\cite{Gehrmann-DeRidder:2019ibf}. In Fig.~\ref{fig:dijets} we present a comparison of the theory prediction at fixed order corrected for non-perturbative and electroweak effects, 
to the CMS dijet 8 TeV triple differential dataset, measured as a function of the dijet average-$p_T$, the rapidity separation  $y^*=1/2|y_1-y_2|$ and boost $y^b=1/2|y_1+y_2|$ of the dijet system. We observe that the NNLO prediction
changes both the shape and normalization of the NLO result and has significantly reduced theory uncertainties. Overall the agreement with the data is excellent at small $y^b$, but for events with large dijet boost kinematics,
the data sits below the theory prediction.  In this region, which is sensitive to the scattering of large-$x$ parton on a low-$x$ parton, the PDFs suffer from large uncertainties. The results 
suggest that the understanding of the high-$x$ behaviour of the PDFs can be improved upon by including 
this dataset in global PDF determinations. A first study in this direction has been presented in~\cite{AbdulKhalek:2020jut}.

\begin{figure}[!b]
\centering
\includegraphics[scale=0.3]{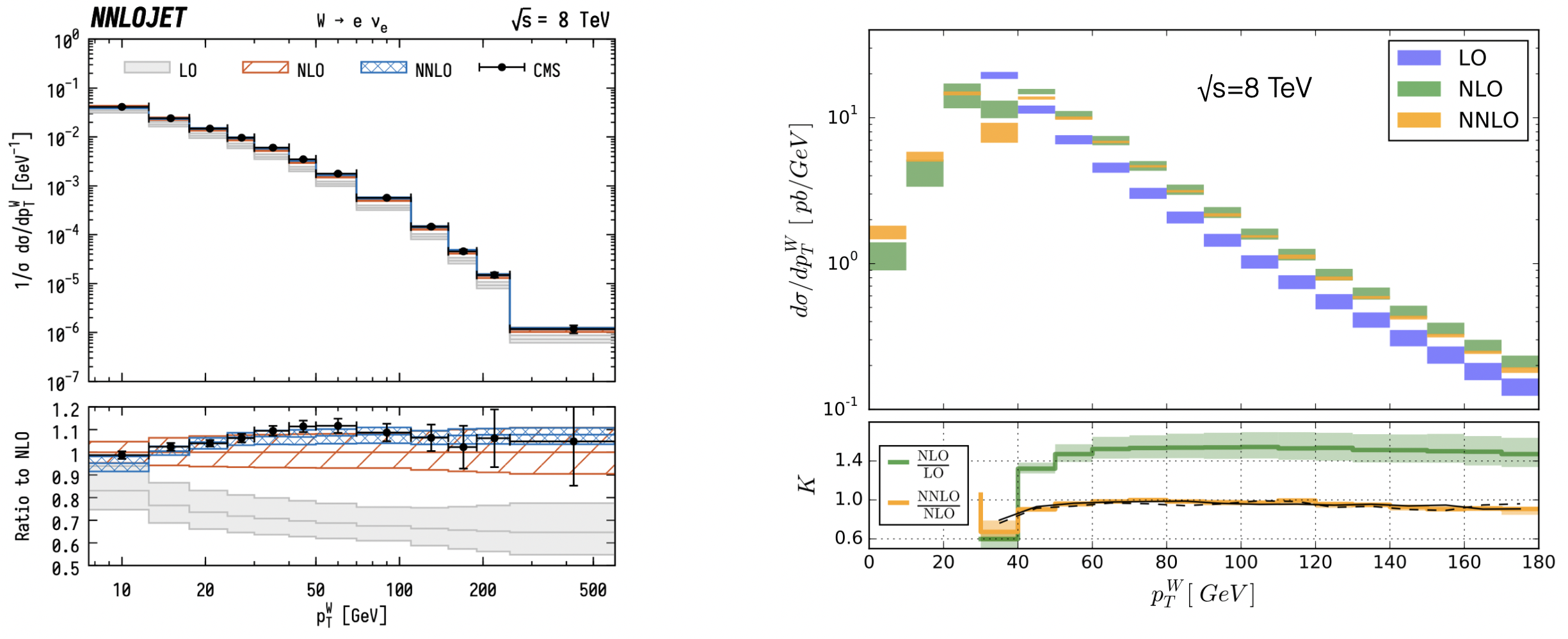}
\caption{$p_{T,W}$ distribution calculated at LO, NLO and NNLO presented in the publications~\cite{Gehrmann-DeRidder:2017mvr} (left) (compared compared to CMS data from Ref.~\cite{CMS:2016mwa}) and~\cite{Boughezal:2015dva} (right). 
The shaded bands correspond to the theoretical scale uncertainties in the calculations.}
\label{fig:Wpt}
\end{figure}

\section{Vector boson plus jet production}
The production of electroweak gauge bosons $(V = \gamma, Z, W±)$ with final state jets has a large production cross section and a clean experimental signature at the LHC. By requiring an additional jet in the final state, we gain sensitivity
to perturbative QCD effects and to the transverse momentum distribution of the gauge boson. The observation of $V$+jet production represents a clean testing ground for perturbative QCD being sensitive to the strong coupling 
constant $\alpha_s$ and the gluon PDF at the lowest order in perturbation theory. 

Perturbative calculations for $W$+jet and $Z$+jet production at NNLO by two independent groups have been presented in~\cite{Boughezal:2015dva,Gehrmann-DeRidder:2017mvr}
and~\cite{Gehrmann-DeRidder:2015wbt,Boughezal:2015ded,Gehrmann-DeRidder:2016cdi} respectively. In Fig.~\ref{fig:Wpt} we present the results for the transverse momentum spectrum of the W-boson at LO, NLO and NNLO in perturbation theory. 
We observe that the NLO corrections are between 10–40\% with residual theoretical scale uncertainties at the level of $\pm10\%$.  The NNLO corrections in turn are at the 5\% level, changing the shape of the 
NLO result and improving the description of the data. Overall the NNLO scale uncertainties are at the $\pm2\%$ level and overlap with the NLO result, indicating a good convergence of the perturbative expansion for this observable. 
 
Results for $\gamma$+jet production at NNLO have been obtained using a smooth-cone photon isolation prescription in~\cite{Campbell:2017dqk} and using a smooth and hybrid photon isolation prescriptions in~\cite{Chen:2019zmr}. We observe
agreement between the two calculation when using same input settings and photon isolation prescriptions~\cite{Chen:2019zmr}. Similarly to the other $V$+jet processes, it is observed a good convergence and stability 
of the perturbative expansion in the NNLO prediction combined with a photon isolation treatment that follows closely the procedure used in experiments. A detailed study of the scale uncertainty
of the theoretical prediction for $\gamma$+jet production has been presented in~\cite{Chen:2019zmr}, where we can observe a dramatic reduction in the scale uncertainties when going from NLO to NNLO. We conclude that all $V$+jet 
processes are known at NNLO in QCD with residual scale uncertainties in the perturbative expansion that are typically at the level of 5\%.

\section{Higgs boson plus jet production}
\begin{figure}[!t]
\centering
\includegraphics[scale=0.3]{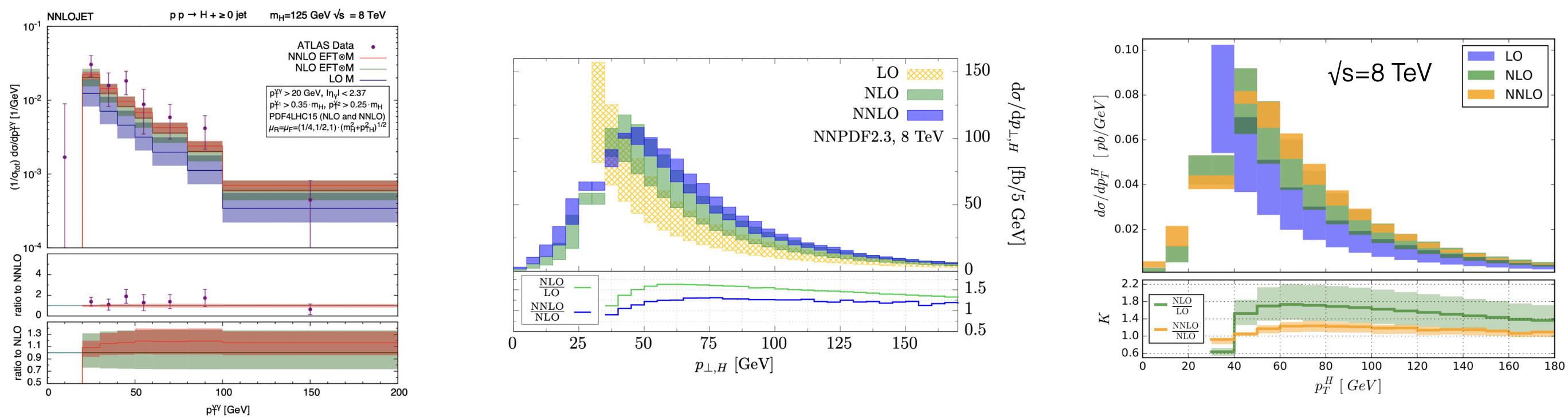}
\caption{Higgs boson transverse momentum distribution calculated at LO, NLO and NNLO in the publications \cite{Chen:2016zka} (left)~\cite{Boughezal:2015dra} (middle),~\cite{Boughezal:2015aha} (right) 
and ATLAS data from ref~\cite{ATLAS:2014yga}. Shaded bands in the plots represent the theoretical scale uncertainty in the calculations.}
\label{fig:hjet}
\end{figure}
The determination of the Higgs boson properties is a central goal at the LHC and HL-LHC, where it is expected that the predictions of the SM will be tested to the five percent level in several production and decay modes. 
For this reason, to improve the modelling of the Higgs kinematics at the LHC, precise predictions for Higgs production in association with jets are needed. The number of jets produced in association with a 
Higgs boson candidate is a very important discriminator between different production modes, and plays a key role in the background rejection for many Higgs boson studies. 

Fixed-order predictions for $H$+jet at NNLO in QCD have been calculated in~\cite{Chen:2016zka,Boughezal:2015dra,Boughezal:2015aha}, in the effective theory approximation, obtained by 
integrating out the top quark. The full top-quark mass dependence for $H$+jet is currently known at NLO only~\cite{Jones:2018hbb}. In Fig.~\ref{fig:hjet} we show results for the Higgs $p_T$-spectrum at NNLO. 
We observe substantially larger radiative corrections with respect to the $V$+jet processes, and a good agreement between the NNLO prediction reweighted by exact top-mass dependence at 
LO and the ATLAS ($H\to \gamma\gamma$) data normalized to the total inclusive Higgs cross section.
 
\begin{figure}[!t]
\centering
\includegraphics[scale=0.32]{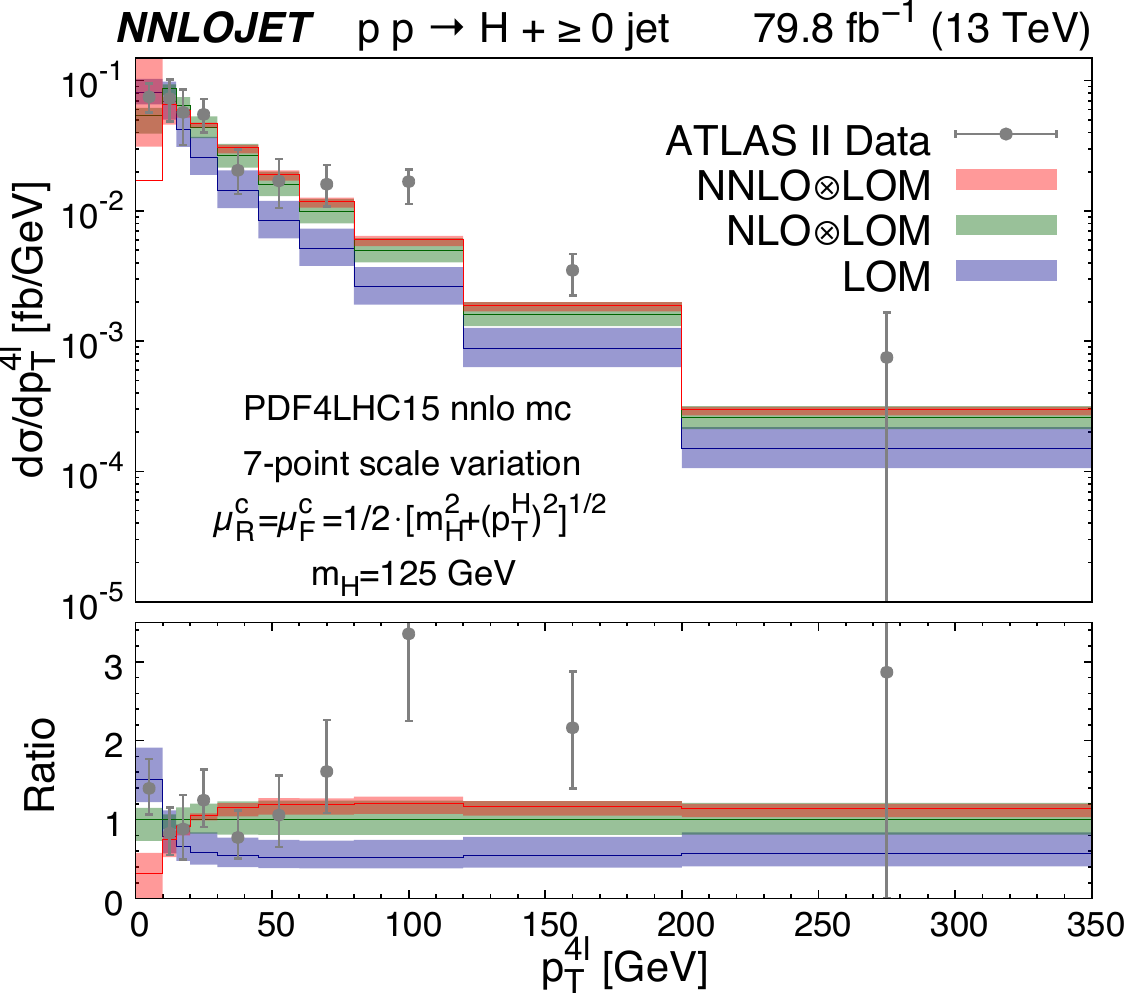}
\includegraphics[scale=0.32]{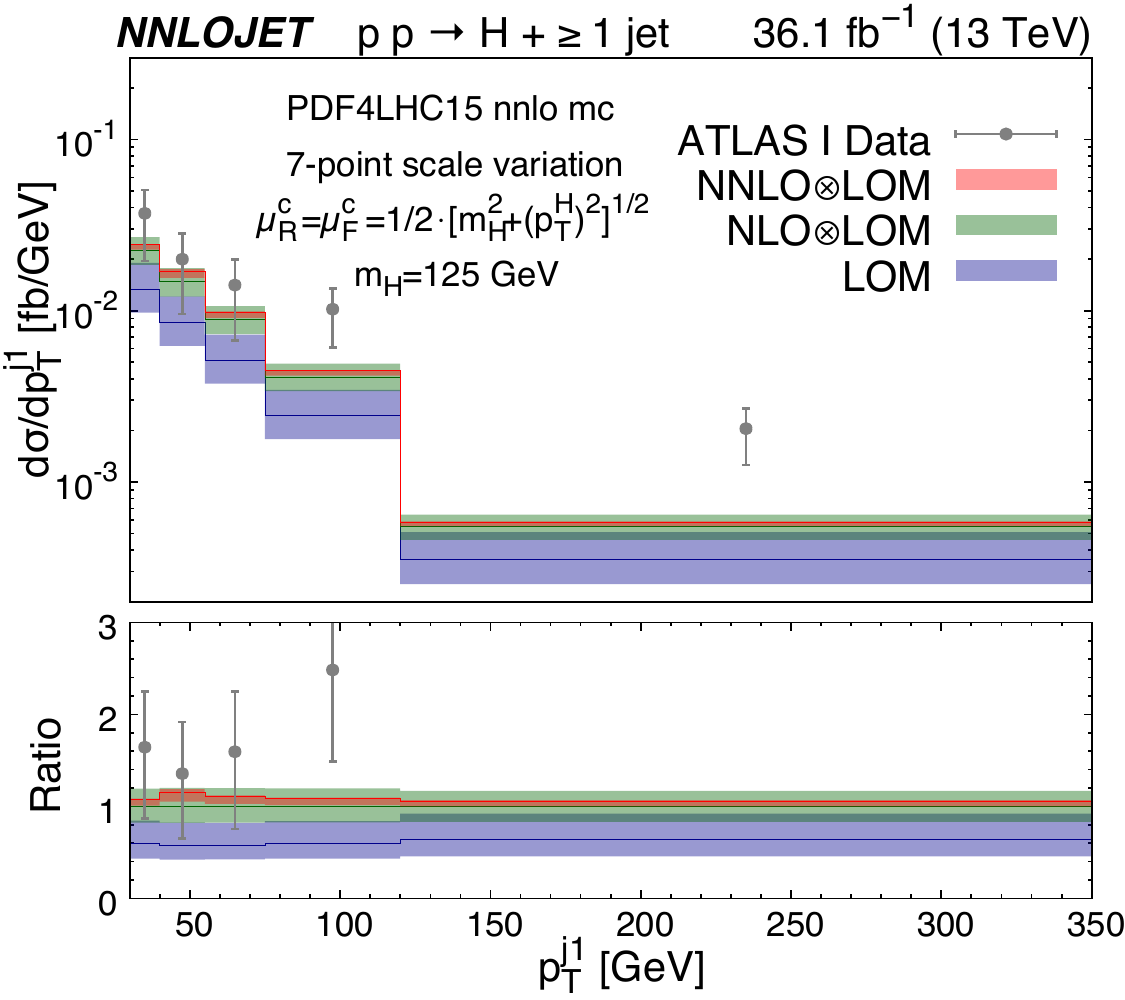}
\includegraphics[scale=0.32]{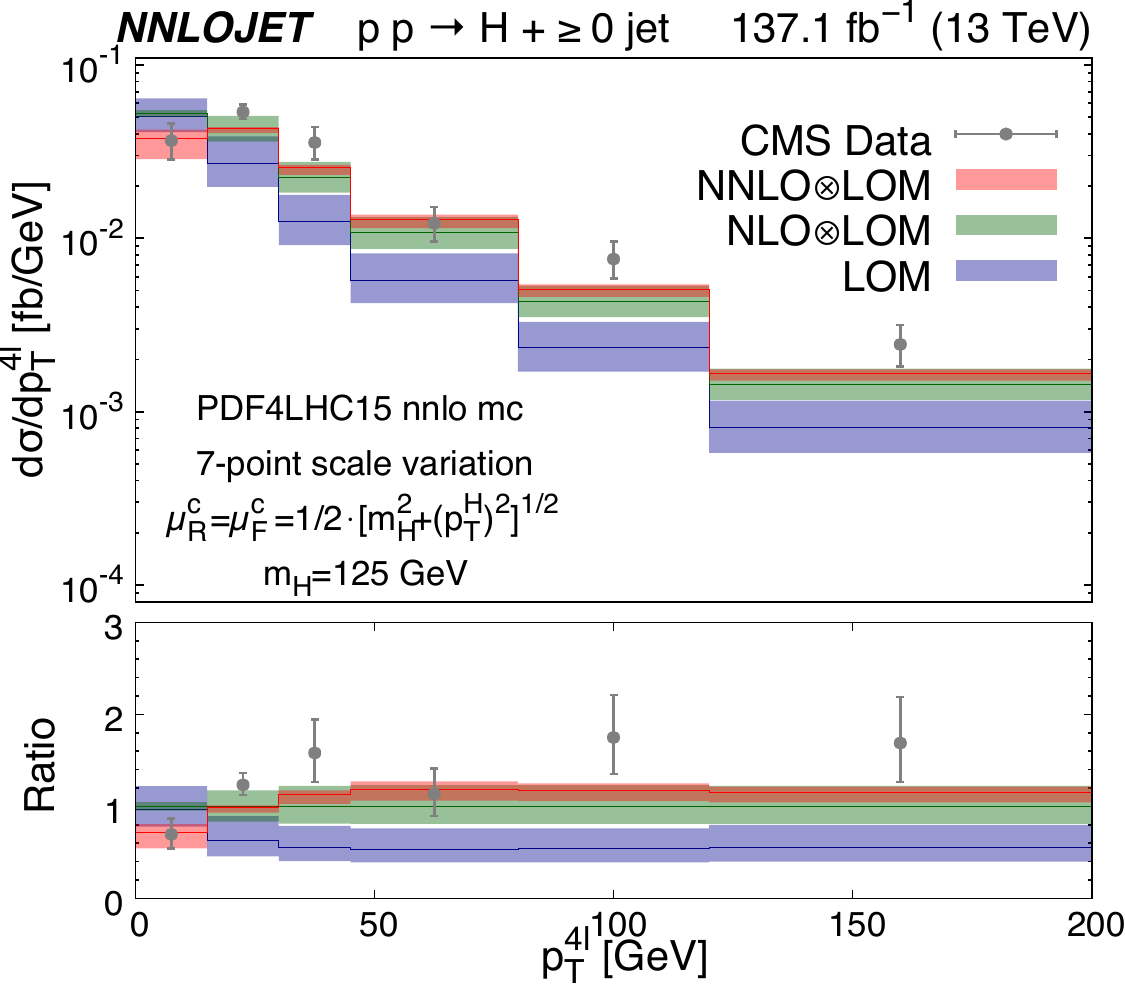}
\includegraphics[scale=0.32]{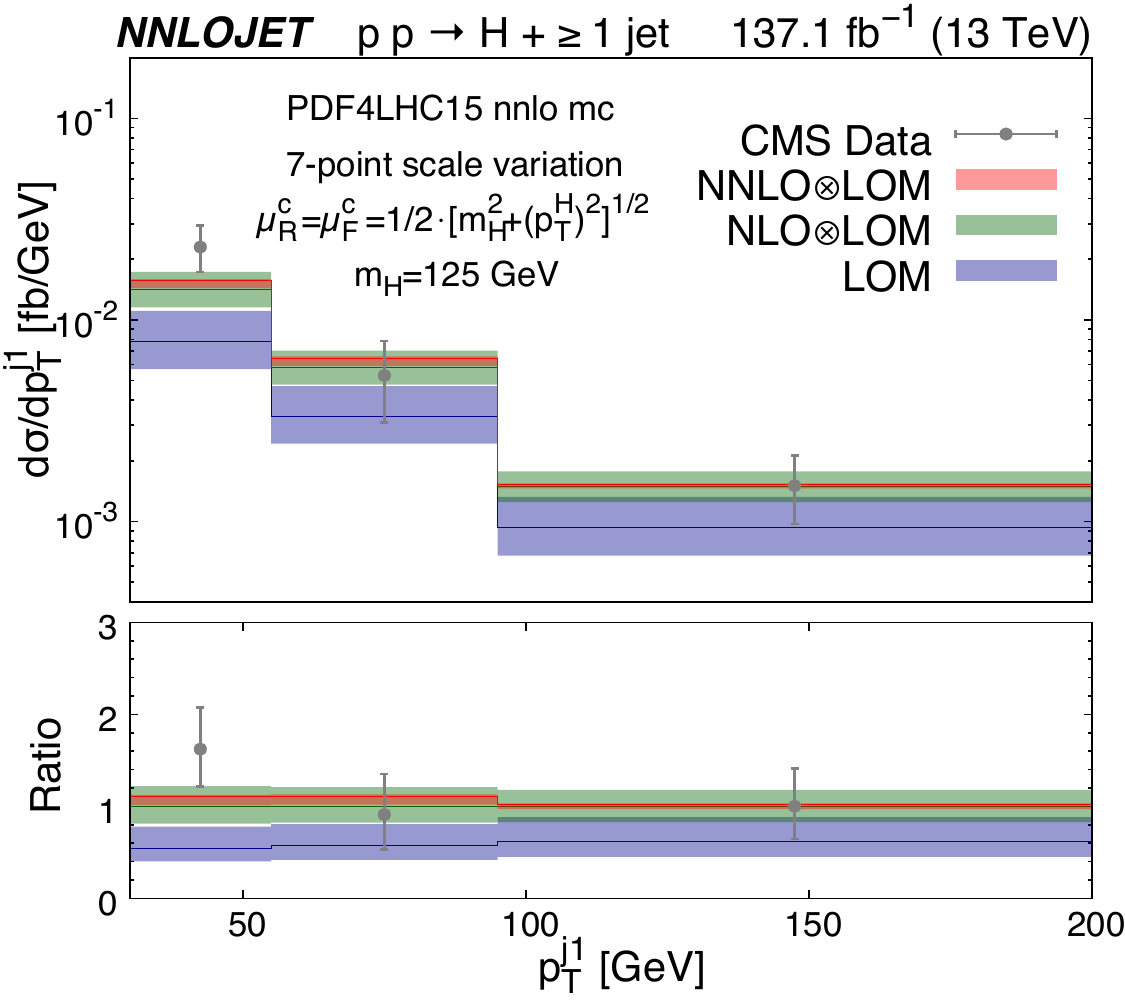}
\caption{Transverse momentum distributions of the Higgs boson and of the leading jet produced in association with a Higgs boson at LO, NLO and NNLO~\cite{Chen:2019wxf} compared 
with ATLAS~\cite{ATLAS:2017qey,ATLAS:2020wny} and CMS~\cite{CMS:2021ugl} data.}
\label{fig:hjet2}
\end{figure}

The four-lepton decay mode of the Higgs boson allows for a clean kinematic reconstruction, thereby enabling precision studies of the Higgs boson properties and of its production dynamics. In Fig.~\ref{fig:hjet2}
we show the NNLO QCD corrections to fiducial cross sections relevant to this decay mode in the gluon-fusion production of a Higgs boson in association with a hadronic jet. In this calculation~\cite{Chen:2019wxf} 
we can take into account fiducial cuts on the Higgs boson decay products as well as on accompanying objects such as hadronic jets. We observe that the NNLO corrections are sizeable and kinematics dependent.
With respect to NLO we observe a substantial reduction of scale uncertainties of the theory prediction to a level of about 10\% in most distributions~\cite{Chen:2019wxf}.

\begin{figure}[!b]
\centering
\includegraphics[scale=0.2]{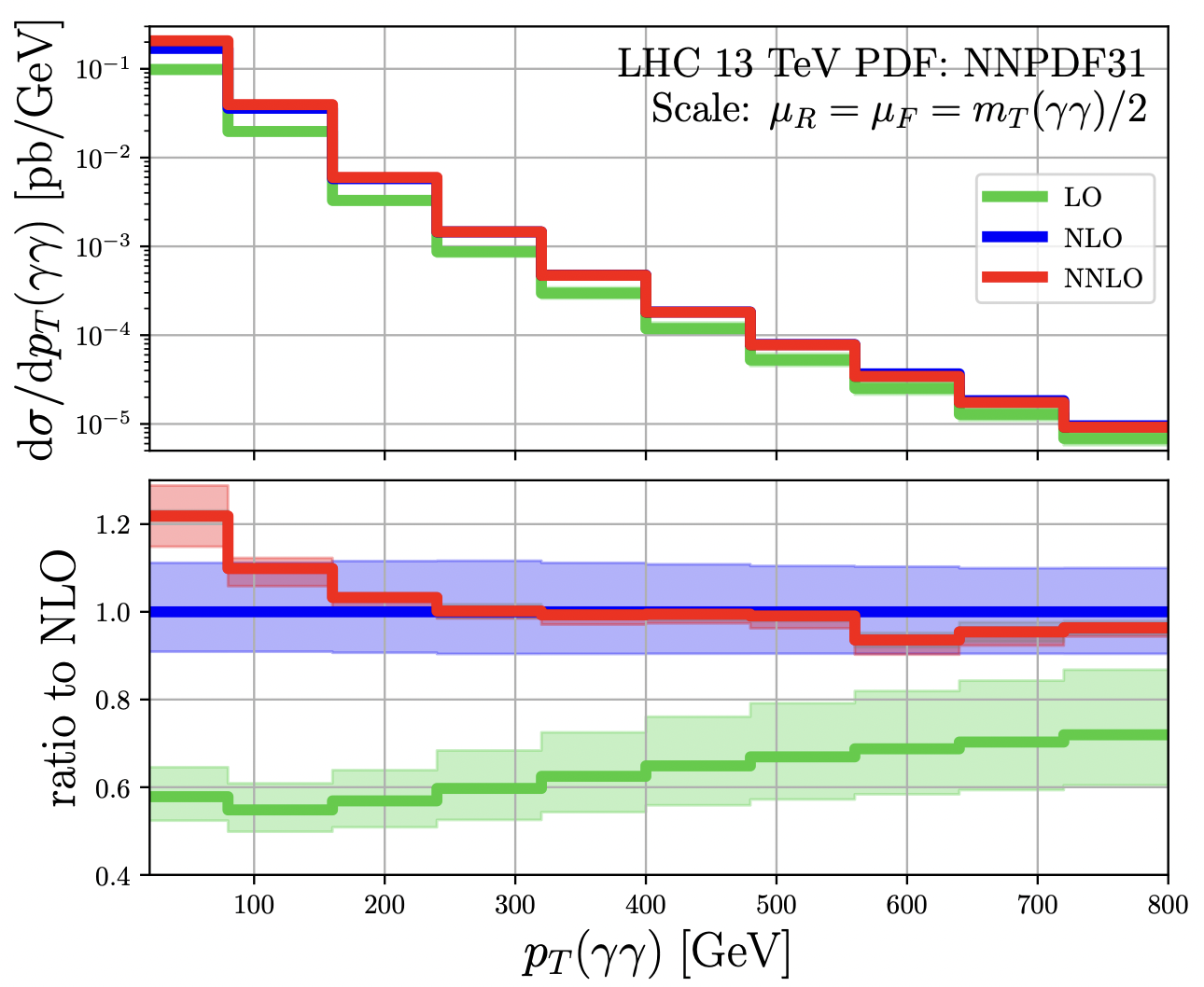}
\caption{Absolute $p_{T}$($\gamma\gamma$) distributions at LO, NLO and NNLO. The colored bands in the plots represent the theoretical scale uncertainty in the calculation~\cite{Chawdhry:2021hkp}.}
\label{fig:aaj}
\end{figure}

\section{Beyond $2\to2$ processes}
\begin{figure}[!t]
\centering
\includegraphics[scale=0.32]{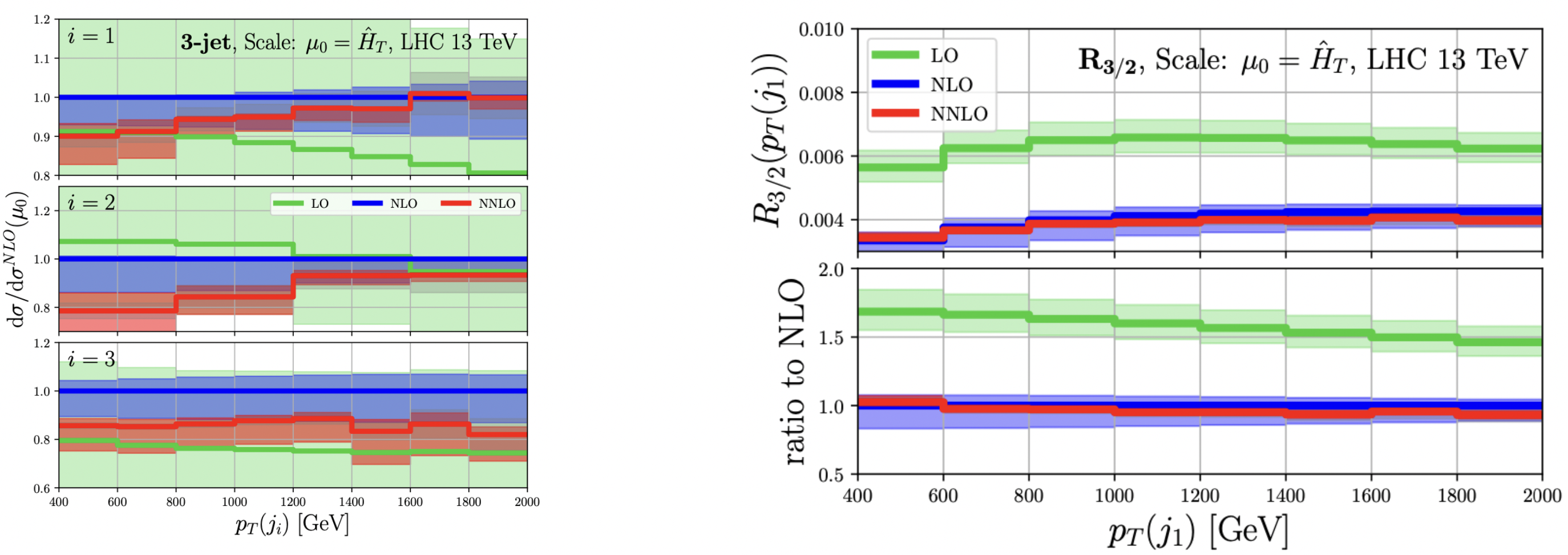}
\caption{Transverse momentum distribution of the three leading jets for three jet production at LO, NLO and NNLO (left). The three jet to two jet ratio as a function of the leading jet $p_T$ (right).
The colored bands in the plots represent the theoretical scale uncertainty in the calculations~\cite{Czakon:2021mjy}.}
\label{fig:3jet}
\end{figure}
In this section we present recent results on the calculation of NNLO jet cross sections for processes beyond $2\to2$ scattering. These processes are the current frontier of research, for which the required two-loop
5-point amplitudes have been only recently derived. In particular, two loop amplitudes for diphoton+jet production and 3-jet production have been derived in~\cite{Chawdhry:2021mkw} and~\cite{Chicherin:2020oor,Abreu:2021oya}. 

The $\gamma\gamma$+jet NNLO calculation presented in~\cite{Chawdhry:2021hkp} represents the first NNLO-accurate prediction for the transverse momentum distribution of the diphoton system. This observable represents the main 
background for Higgs production at high-$p_{T}$, and is relevant for dedicated measurements of diphoton production. In Fig~\ref{fig:aaj} we show the diphoton $p_{T}$($\gamma\gamma$) spectrum in fixed-order perturbation theory.
We can observe that NNLO is the first order where the perturbative expansion converges with overlapping scale uncertainty bands between two consecutive orders. The scale uncertainty in the observable is at the 1$\sim$2\% level
across most of the distribution.

In Fig.\ref{fig:3jet} we present distributions for three jet production at NNLO in QCD calculated in~\cite{Czakon:2021mjy}. This process has been studied in great detail by the experimental collaborations
with the aim of providing a unique testing of perturbative QCD, by comparing theory predictions with collider jet data. However, current experimental analysis of this process have been 
limited by the scale uncertainties of the NLO calculation. In this regard it is expected that the NNLO calculation will allow for improved predictions for jet transverse momenta, angular correlations and event-shape observables
with substantially reduced scale uncertainties. 

In Fig.\ref{fig:3jet} (left) we can observe that the NNLO corrections are of the of the order of -15\% at low-$p_T$ and increase steadily at high-$p_T$ for the first and second jet-$p_T$ spectrum. 
The 3rd jet-$p_{T}$ distribution is well-behaved: it has a flat NNLO corrections and fairly symmetric uncertainty bands at both NLO and NNLO. In the same figure on the right it is shown the
three-jet to two-jet ratio as a function of the $p_T$ of the leading jet. We can observe that the prediction for the ratio is stabilised at NNLO with scale uncertainties at the 3\% level. 

\section{Conclusions}
Calculations of NNLO QCD corrections for many important $2\to2$ and $2\to3$ processes with jets in the final state at the LHC are now available and were reviewed in this presentation. The impressive progress in this field
has been made possible due to the development of subtraction methods to treat infrared divergences at this order, and due to many advances in the calculation of multi-loop integrals and amplitudes.
Given the amount of experimental data due to be collected during the LHC and HL-LHC runs, we anticipate that these results will allow us to perform physics analysis with LHC data 
at a new level of precision.


\begin{thebibliography}{99}
\bibitem{Currie:2016bfm}
J.~Currie, E.~W.~N.~Glover and J.~Pires,
Phys. Rev. Lett. \textbf{118} (2017) no.7, 072002
doi:10.1103/PhysRevLett.118.072002
[arXiv:1611.01460 [hep-ph]].

\bibitem{Currie:2017eqf}
J.~Currie, A.~Gehrmann-De Ridder, T.~Gehrmann, E.~W.~N.~Glover, A.~Huss and J.~Pires,
Phys. Rev. Lett. \textbf{119} (2017) no.15, 152001
doi:10.1103/PhysRevLett.119.152001
[arXiv:1705.10271 [hep-ph]].

\bibitem{Gehrmann-DeRidder:2019ibf}
A.~Gehrmann-De Ridder, T.~Gehrmann, E.~W.~N.~Glover, A.~Huss and J.~Pires,
Phys. Rev. Lett. \textbf{123} (2019) no.10, 102001
doi:10.1103/PhysRevLett.123.102001
[arXiv:1905.09047 [hep-ph]].

\bibitem{Czakon:2019tmo}
M.~Czakon, A.~van Hameren, A.~Mitov and R.~Poncelet,
JHEP \textbf{10} (2019), 262
doi:10.1007/JHEP10(2019)262
[arXiv:1907.12911 [hep-ph]].

\bibitem{AbdulKhalek:2020jut}
R.~Abdul Khalek, S.~Forte, T.~Gehrmann, A.~Gehrmann-De Ridder, T.~Giani, N.~Glover, A.~Huss, E.~R.~Nocera, J.~Pires and J.~Rojo, \textit{et al.}
Eur. Phys. J. C \textbf{80} (2020) no.8, 797
doi:10.1140/epjc/s10052-020-8328-5
[arXiv:2005.11327 [hep-ph]].

\bibitem{ATLAS:2017ble}
M.~Aaboud \textit{et al.} [ATLAS],
JHEP \textbf{05} (2018), 195
doi:10.1007/JHEP05(2018)195
[arXiv:1711.02692 [hep-ex]].

\bibitem{CMS:2021yzl}
A.~Tumasyan \textit{et al.} [CMS],
[arXiv:2111.10431 [hep-ex]].

\bibitem{Gehrmann-DeRidder:2017mvr}
A.~Gehrmann-De Ridder, T.~Gehrmann, E.~W.~N.~Glover, A.~Huss and D.~M.~Walker,
Phys. Rev. Lett. \textbf{120} (2018) no.12, 122001
doi:10.1103/PhysRevLett.120.122001
[arXiv:1712.07543 [hep-ph]].

\bibitem{CMS:2016mwa}
V.~Khachatryan \textit{et al.} [CMS],
JHEP \textbf{02} (2017), 096
doi:10.1007/JHEP02(2017)096
[arXiv:1606.05864 [hep-ex]].

\bibitem{Boughezal:2015dva}
R.~Boughezal, C.~Focke, X.~Liu and F.~Petriello,
Phys. Rev. Lett. \textbf{115} (2015) no.6, 062002
doi:10.1103/PhysRevLett.115.062002
[arXiv:1504.02131 [hep-ph]].

\bibitem{Gehrmann-DeRidder:2015wbt}
A.~Gehrmann-De Ridder, T.~Gehrmann, E.~W.~N.~Glover, A.~Huss and T.~A.~Morgan,
Phys. Rev. Lett. \textbf{117} (2016) no.2, 022001
doi:10.1103/PhysRevLett.117.022001
[arXiv:1507.02850 [hep-ph]].

\bibitem{Boughezal:2015ded}
R.~Boughezal, J.~M.~Campbell, R.~K.~Ellis, C.~Focke, W.~T.~Giele, X.~Liu and F.~Petriello,
Phys. Rev. Lett. \textbf{116} (2016) no.15, 152001
doi:10.1103/PhysRevLett.116.152001
[arXiv:1512.01291 [hep-ph]].

\bibitem{Gehrmann-DeRidder:2016cdi}
A.~Gehrmann-De Ridder, T.~Gehrmann, E.~W.~N.~Glover, A.~Huss and T.~A.~Morgan,
JHEP \textbf{07} (2016), 133
doi:10.1007/JHEP07(2016)133
[arXiv:1605.04295 [hep-ph]].

\bibitem{Campbell:2017dqk}
J.~M.~Campbell, R.~K.~Ellis and C.~Williams,
Phys. Rev. D \textbf{96} (2017) no.1, 014037
doi:10.1103/PhysRevD.96.014037
[arXiv:1703.10109 [hep-ph]].

\bibitem{Chen:2019zmr}
X.~Chen, T.~Gehrmann, N.~Glover, M.~H\"ofer and A.~Huss,
JHEP \textbf{04} (2020), 166
doi:10.1007/JHEP04(2020)166
[arXiv:1904.01044 [hep-ph]].

\bibitem{Chen:2016zka}
X.~Chen, J.~Cruz-Martinez, T.~Gehrmann, E.~W.~N.~Glover and M.~Jaquier,
JHEP \textbf{10} (2016), 066
doi:10.1007/JHEP10(2016)066
[arXiv:1607.08817 [hep-ph]].

\bibitem{Boughezal:2015dra}
R.~Boughezal, F.~Caola, K.~Melnikov, F.~Petriello and M.~Schulze,
Phys. Rev. Lett. \textbf{115} (2015) no.8, 082003
doi:10.1103/PhysRevLett.115.082003
[arXiv:1504.07922 [hep-ph]].

\bibitem{Boughezal:2015aha}
R.~Boughezal, C.~Focke, W.~Giele, X.~Liu and F.~Petriello,
Phys. Lett. B \textbf{748} (2015), 5-8
doi:10.1016/j.physletb.2015.06.055
[arXiv:1505.03893 [hep-ph]].

\bibitem{Jones:2018hbb}
S.~P.~Jones, M.~Kerner and G.~Luisoni,
Phys. Rev. Lett. \textbf{120} (2018) no.16, 162001
doi:10.1103/PhysRevLett.120.162001
[arXiv:1802.00349 [hep-ph]].

\bibitem{ATLAS:2014yga}
G.~Aad \textit{et al.} [ATLAS],
JHEP \textbf{09} (2014), 112
doi:10.1007/JHEP09(2014)112
[arXiv:1407.4222 [hep-ex]].

\bibitem{Chen:2019wxf}
X.~Chen, T.~Gehrmann, E.~W.~N.~Glover and A.~Huss,
JHEP \textbf{07} (2019), 052
doi:10.1007/JHEP07(2019)052
[arXiv:1905.13738 [hep-ph]].

\bibitem{ATLAS:2017qey}
M.~Aaboud \textit{et al.} [ATLAS],
JHEP \textbf{10} (2017), 132
doi:10.1007/JHEP10(2017)132
[arXiv:1708.02810 [hep-ex]].

\bibitem{ATLAS:2020wny}
G.~Aad \textit{et al.} [ATLAS],
Eur. Phys. J. C \textbf{80} (2020) no.10, 942
doi:10.1140/epjc/s10052-020-8223-0
[arXiv:2004.03969 [hep-ex]].

\bibitem{CMS:2021ugl}
A.~M.~Sirunyan \textit{et al.} [CMS],
Eur. Phys. J. C \textbf{81} (2021) no.6, 488
doi:10.1140/epjc/s10052-021-09200-x
[arXiv:2103.04956 [hep-ex]].

\bibitem{Chawdhry:2021mkw}
H.~A.~Chawdhry, M.~Czakon, A.~Mitov and R.~Poncelet,
JHEP \textbf{07} (2021), 164
doi:10.1007/JHEP07(2021)164
[arXiv:2103.04319 [hep-ph]].

\bibitem{Chicherin:2020oor}
D.~Chicherin and V.~Sotnikov,
JHEP \textbf{20} (2020), 167
doi:10.1007/JHEP12(2020)167
[arXiv:2009.07803 [hep-ph]].

\bibitem{Abreu:2021oya}
S.~Abreu, F.~F.~Cordero, H.~Ita, B.~Page and V.~Sotnikov,
JHEP \textbf{07} (2021), 095
doi:10.1007/JHEP07(2021)095
[arXiv:2102.13609 [hep-ph]].

\bibitem{Chawdhry:2021hkp}
H.~A.~Chawdhry, M.~Czakon, A.~Mitov and R.~Poncelet,
JHEP \textbf{09} (2021), 093
doi:10.1007/JHEP09(2021)093
[arXiv:2105.06940 [hep-ph]].

\bibitem{Czakon:2021mjy}
M.~Czakon, A.~Mitov and R.~Poncelet,
Phys. Rev. Lett. \textbf{127} (2021) no.15, 152001
doi:10.1103/PhysRevLett.127.152001
[arXiv:2106.05331 [hep-ph]].


\end{thebibliography}
\end{document}